# An Instrument for Precision Controlled Radiation Exposures, Charged Beam Profile Measurement, and Real-time Fluence Monitoring Beyond $10^{16}$ $n_{eq}/cm^2$

M.R. Hoeferkamp, J.S.T. Wickramasinghe, A. Grummer, I. Rajen, S. Seidel
Department of Physics and Astronomy, University of New Mexico,
210 Yale Blvd. NE, Albuquerque, NM 87106 USA

**Abstract**

An instrument has been developed for precision controlled exposures of electronic devices and material samples in particle beams. The instrument provides simultaneously a real time record of the profile of the beam and the fluence received. The system is capable of treating devices with dimensional scales ranging from millimeters to extended objects of cross sections measured in tens of square centimeters. The instrument has been demonstrated to operate effectively in integrated fluences ranging up to a few times $10^{16}$ 1-MeV-neutron-equivalent/cm$^2$ ($n_{eq}$). The positioner portion of the system comprises a set of remotely controllable sample holders incorporating cooling and interfaces for sample power and readout, all constructed from low activation technologies. The monitoring component of the system samples the current or voltage of radiation tolerant silicon diodes placed directly in the path of the beam.

**Introduction**

A custom instrument has been developed to facilitate remote control and monitoring of samples in a particle beam. One purpose of this article is to describe the device in sufficient detail to permit replication of it. A second goal is to describe the branch points in the design process, so that an investigator with goals that are related but not identical could build upon this experience to optimize differently.

This is a portable instrument that provides remote control of the position and environment of experimental samples in a radiation zone for use by experimenters who are outside of the radiation enclosure. Being portable and robust for transportation, this instrument can be used at a variety of beamlines worldwide. The capability of real-time measurement of beam profile and fluence can be used both to prepare the beam prior to exposure and to monitor the beam during exposure. On the order of one hundred distinct samples can be served by this instrument simultaneously.

**The Positioners**

The horizontal positioning system is shown in the upper half of Figure 1. Several translation stages are mounted to an acrylic platform of dimensions 51 cm × 86 cm × 1.3 cm. The translation arms are perpendicular to the axis to which the beam is aligned. Upon each stage is affixed one sample box using non-activating (plastic) screws except at one position, where a metal screw is required for rigidity. Each stage provides a 10 cm travel distance. Translation is achieved by pneumatic pressure applied differentially, through a manifold (lower half of Figure 1) to the near or far end of a PVC cylinder of length 14 cm and diameter 5.7 cm, displacing a PVC piston internal to the cylinder and attached to its stage by a PVC rod. The exposed and the internal O-rings of the cylinder are EPDM for which the elastic seal integrity has been demonstrated up to $10^6$ Gy or $10^8$ rad. The two ends of each pneumatic tube are attached to 6.4 mm plastic tubing that runs 30 m to the manifold where the application of a simple hand-operated air pump translates a selected sample remotely and practically without delay.

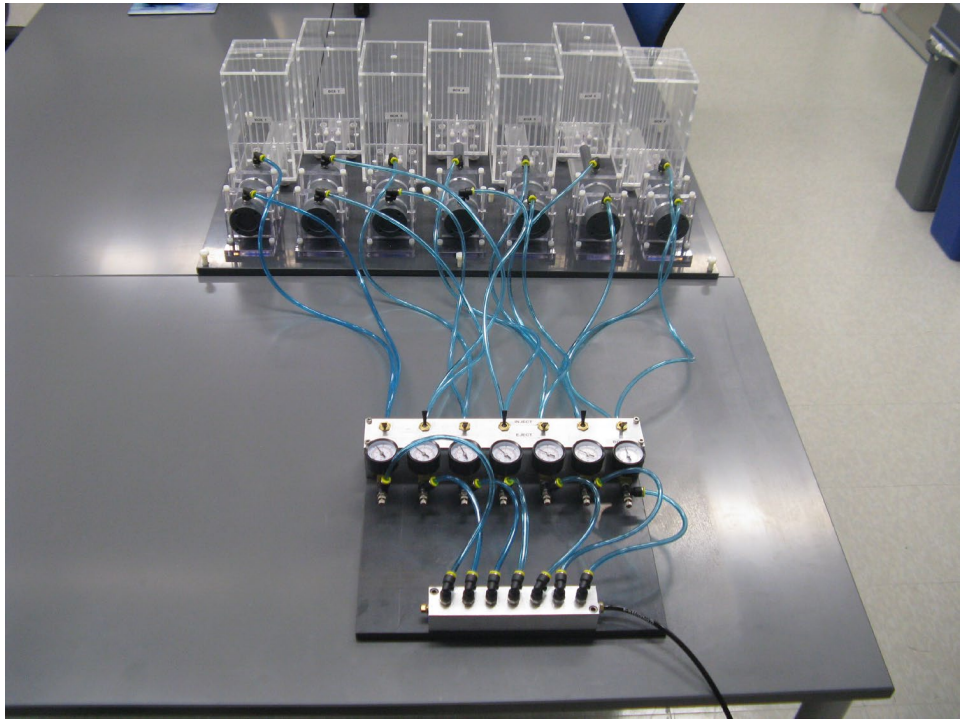

Figure 1: The positioning system, showing seven horizontal translation stages with attached sample boxes and a manifold that allows differential air pressure to be applied selectively to either end of each translation piston. During operation, the actual length of the plastic tubing between manifold and stages would be much greater - typically tens of meters, in order to allow personnel to operate the manifold outside of a radiation zone in which the translation stages are enclosed.

A typical sample box (see Figure 2) has dimensions 13.5 cm × 11.4 cm × 21.9 cm and is fabricated from acrylic and assembled with plastic screws. The beam enters and exits each box sequentially through a Kapton-covered square aperture of dimensions 2.5 cm × 2.5 cm. Boxes have lids through which power, readout, and cooling gas flow connections can be made. Precision machined slots at 1 cm intervals on opposing interior walls of the box have depth 2 mm. These hold G-10 boards onto which samples are affixed, typically with Kapton tape. Using these precision slots, samples whose dimension is small compared to the width of the beam are positioned perpendicular to the beam.

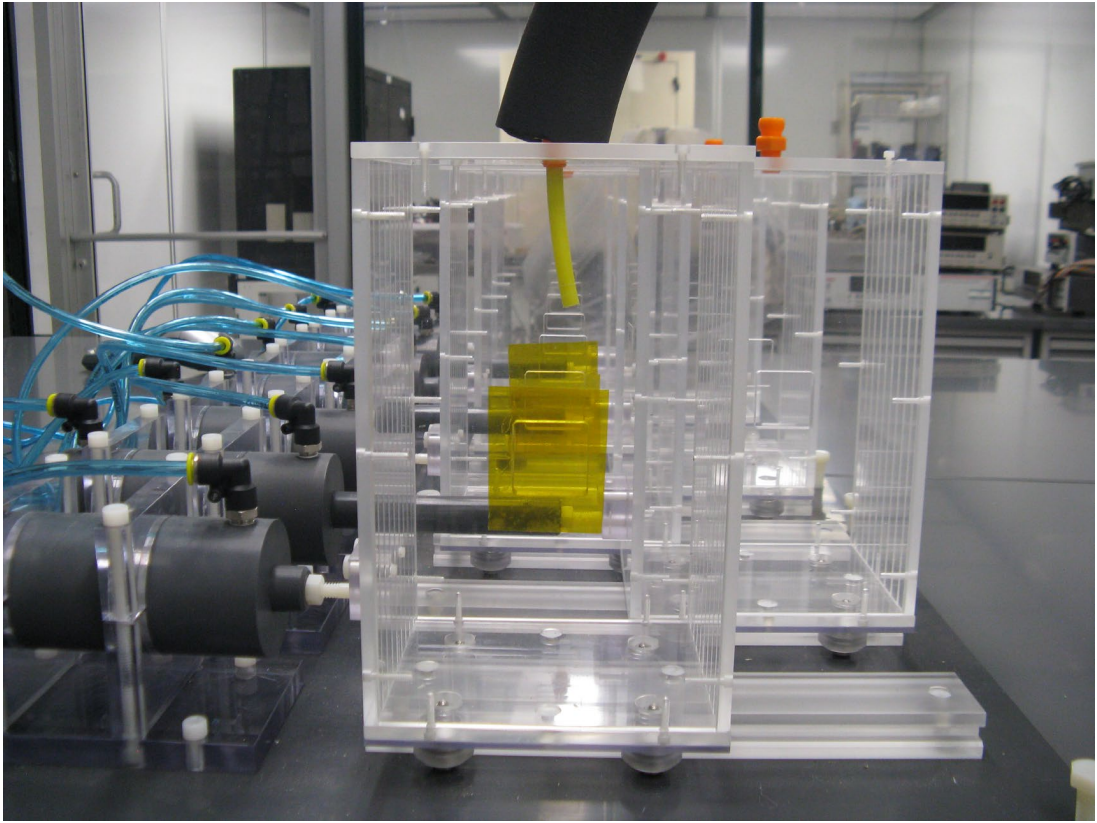

Figure 2: A sample box with gas flow port in the lid.

Samples significantly larger than the beam can be oriented at an angle of 60° with respect to the beam, to reduce their effective cross section by half and thereby permit more uniform exposure in one of their dimensions; for these samples, larger G-10 boards are held in position by slotted inserts positioned at the floor of their box (see Figure 3).

The positioning system also includes a stage that combines vertical and horizontal translation (Figure 4). The vertical translation system has a resolution of 1 μm and can scan through a vertical distance of 12 cm. A separate horizontal stabilizing rod that slides through a linear sleeve bearing is needed to achieve stability and repeatability of the sample box placement and scanning motion. The electronics that control the motor that activates the vertical motion are protected by a lead shield.

The number of sample boxes that can be included is limited only by available space in the beamline.

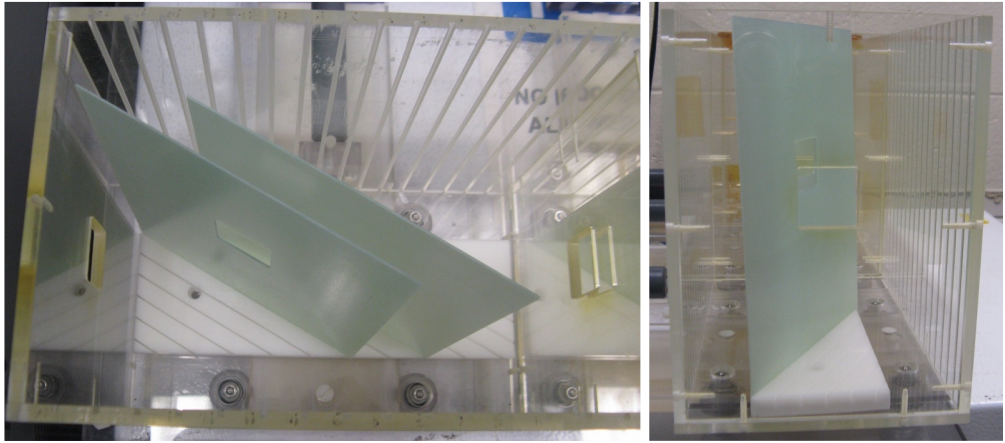

Figure 3: (left) Top view and (right) side view of the removable slotted insert that can be positioned at the base of a sample box to allow the sample to be oriented at an angle with respect to the beam axis.

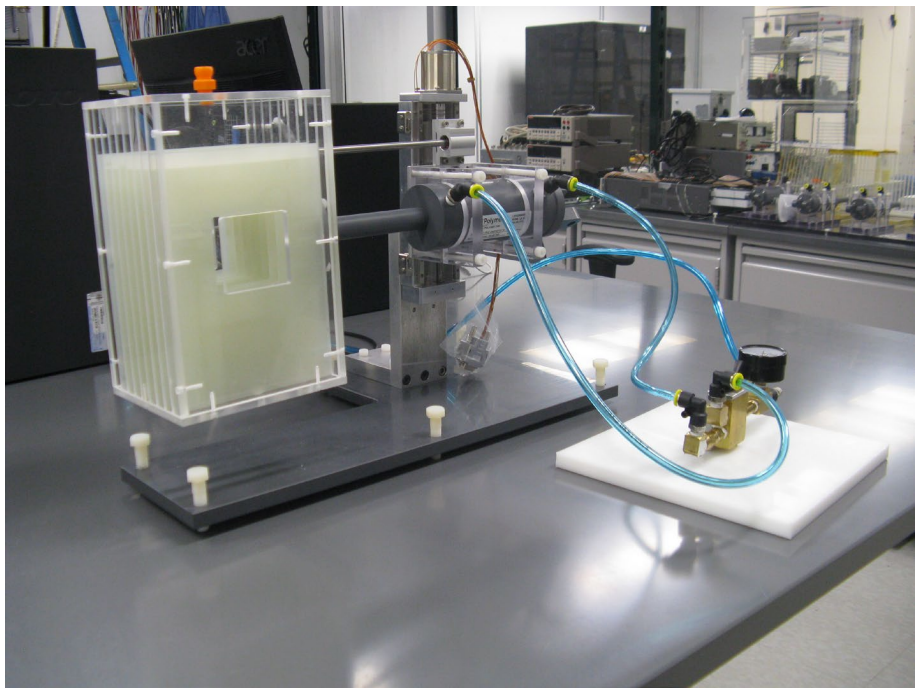

Figure 4: The combined vertical-horizontal translation stage.

**The Beam Monitor**

The beam profile and fluence are measured in real time at one or more points by custom electronics boards. Each of these boards carries a rectangular array of diodes through which the beam passes. Two diode technologies were investigated; both were found to be suitable for particular goals of the experimenter. Both are described here. In both cases, the multiple identical arrays are operated simultaneously in the beam, typically with one placed at each end of the stack of devices under test, permitting some redundancy as well as interpolation of fluences in the spatial interval between them.

*P-i-n diodes*

P-i-n diodes of the type BPW34F by OSRAM have the property that their forward voltage increases linearly with fluence when supplied by a constant forward current. For current amplitude 1 mA and pulse width 50 ms or less, the linear regime in forward voltage versus fluence spans the range from approximately $2 \times 10^{12}$ $n_{eq}/cm^2$ to approximately $4 \times 10^{15}$ $n_{eq}/cm^2$ [1]. The default configuration is a 7×7 array of 49 diodes on a pitch of 3.8 mm, covering an area of 2.5 square centimeters. The diodes are assembled into the array (see Figure 2 in [2]) using standard lead-free solder.

*3D diodes*

Diodes of the 3D geometry, like other silicon sensors, manifest a linear relationship between their leakage current and hadronic fluence given by $I_{leak} = \alpha V \phi_{neq}$, where $\alpha$ is the current-related damage coefficient [3], $V$ is the depleted volume, and $\phi_{neq}$ is the 1-MeV-neutron equivalent fluence. A 7×4 array of 28 diodes was constructed (see Figure 5) with pitch 3.8 mm × 7.6 mm. Because this array is constituted by discrete points, the resolution, given by pitch/√12 [4], is 1.1 mm × 2.2 mm. The array incorporates diodes of types 2E and 3E (which have respectively 2 and 3 $n^+$ columns per pixel cell), some with and some without guard rings, as variation among these features was found not to influence the array's operation. The nominal thickness of every diode is 130 microns. These diodes are very uniform in their characteristics, having been taken from a common wafer and pre-screened for similar electrical characteristics. Prior to irradiation, the leakage current and breakdown voltage of every diode in the batch were measured. Those used for this instrument were selected on the basis of their having pre-irradiation leakage current in the range 30 to 300 pA and pre-irradiation breakdown voltage above 100 V. The pre-irradiation depletion voltages of the diodes were also measured prior to assembly, and all diodes showed full depletion below 2 V. The diodes are held in place with silver conductive epoxy and connected to the readout lines of the array with wire bonds.

*Diode array readout*

Each diode array is connected by ribbon cable to a converter box that maps each channel to a coaxial cable.

In the case of the 3D diodes, the coaxial outputs are connected to Keithley 3761 Low Current Multiplexer cards in a Keithley 3706A scanner (Figure 6). A Keithley 6487 picoammeter supplies voltage and measures the current of each diode through the scanner. The measurement is automated with LabView. The diodes are unpowered except during operator-initiated readout. A complete scan of the 28 diodes requires 3.64 seconds.

The mode of operation of the p-i-n diode array has been described previously [3]. A Keithley 2410 sourcemeter sources a pulse of current and reads out the forward voltage across the *pn* junction. The sourcemeter is interfaced to the array through the Keithley 3706 scanner for rapid automated data collection via LabView.

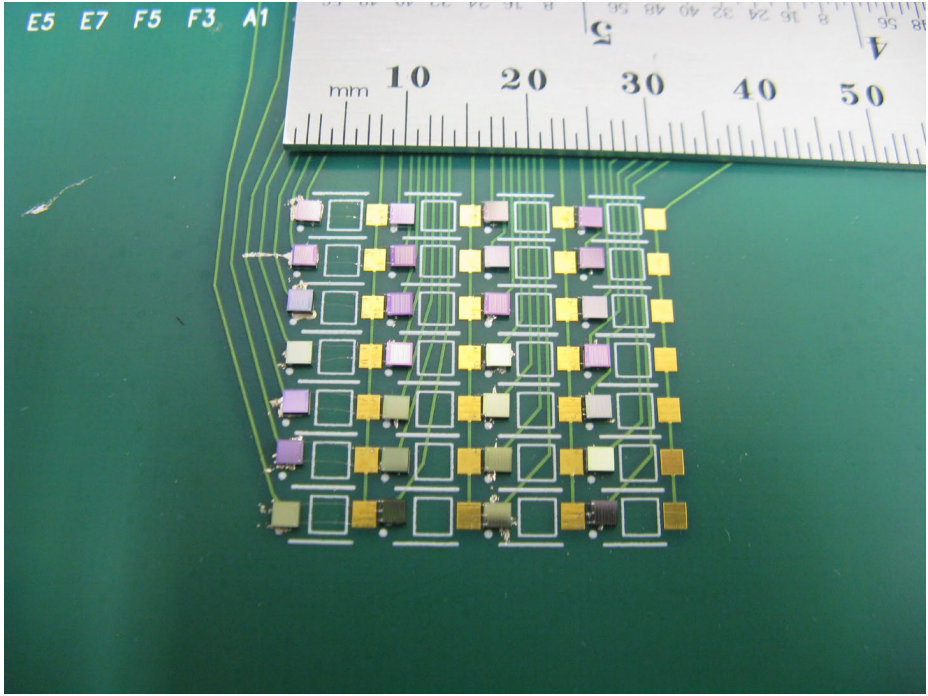

Figure 5: The 3D diode array on a printed circuit board.

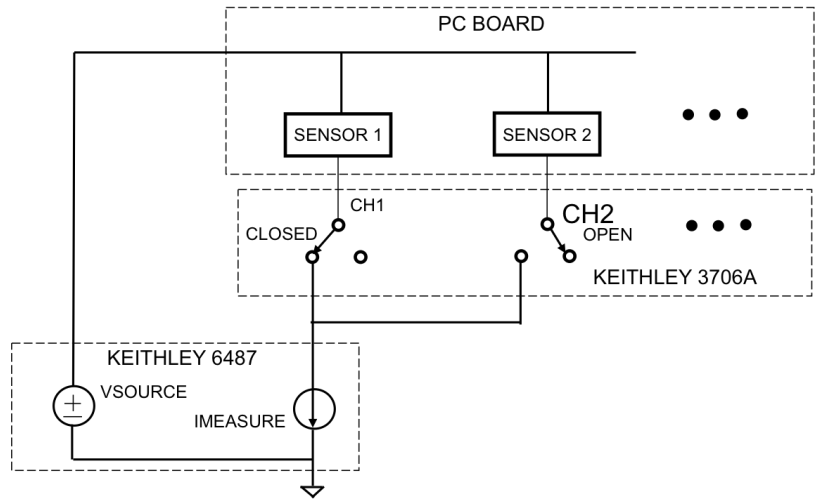

Figure 6: Electrical circuit for the 3D diode array readout.

Figure 7 shows the beam profile obtained using forward voltage data from a p-i-n diode array. An interpolation function supplied by ROOT is used to reconstruct the shape of the beam.

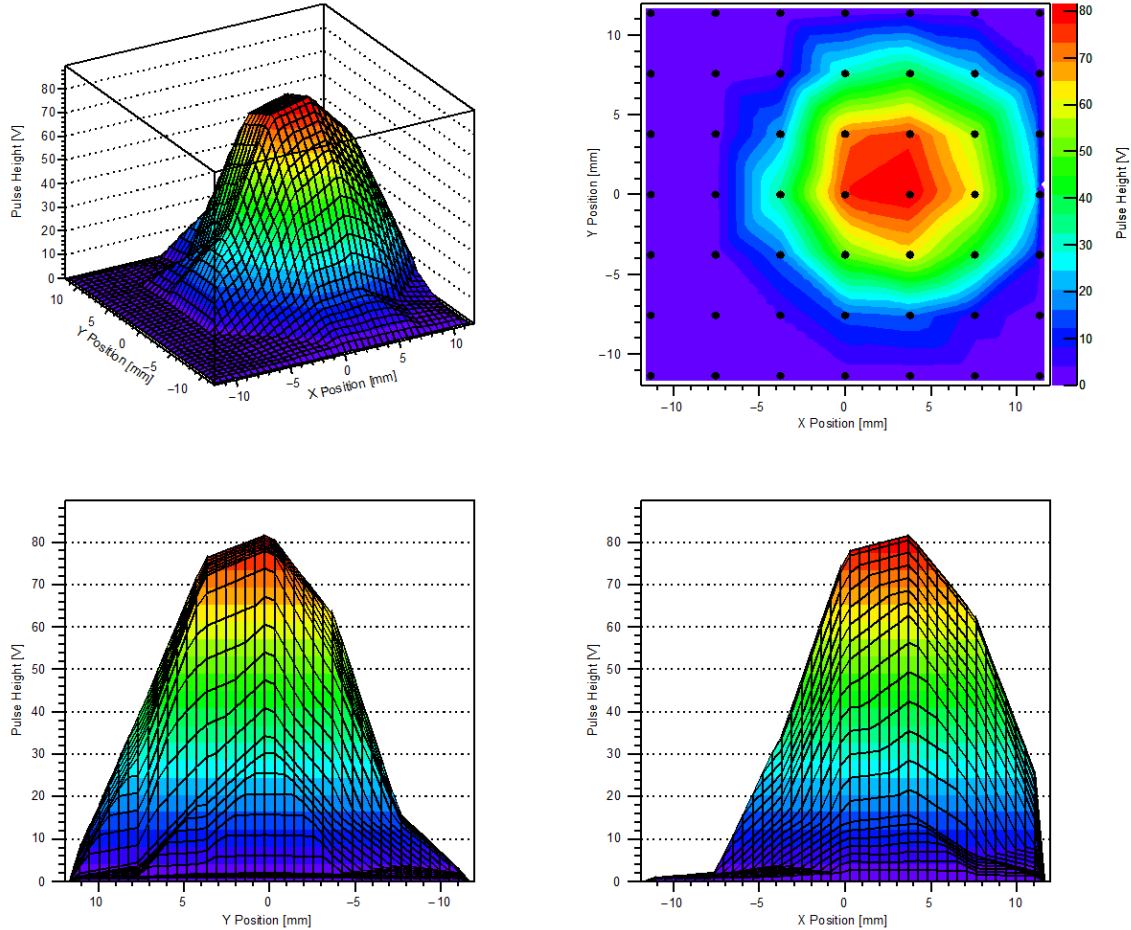

Figure 7: The 800 MeV proton beam profile, imaged using the p-i-n diode array. (Lower left) The recorded pulse height (or voltage) versus position, projected in the y-direction; (Lower right) The recorded pulse height versus position, projected in the x-direction; (Upper right) three-dimensional reconstruction of the pulse height recorded as a function of x- and y-positions; (Upper left) contour plot of pulse height in the x-y plane, with black dots indicating the locations of the centers of the diodes.

The beam profile, recorded with a different beam using leakage current data of 3D diodes, is shown in Figure 8. An interpolation function supplied by ROOT is used to reconstruct the shape of the beam. Three of the twenty-eight 3D diodes are excluded from the analysis due to broken wirebonds. Table 1 summarizes the sources of uncertainty on the 3D leakage current data. An uncertainty of 8.2% is associated with the effects of packaging (epoxy, wirebonds, and light-tight enclosure) upon the 3D diodes in the array; this is obtained by comparing pre-irradiation leakage current data acquired with bare diodes on the probe station to data acquired after assembly of the array in the printed circuit board. The data are transferred to the computer through a 30 m long ribbon cable, which introduces an uncertainty of 11.7%. This uncertainty arises from the data acquisition procedure, which reads out bias current in response to pulsed bias voltage; RLC characteristics of the cable modify the pulse shape. This uncertainty was measured by comparing data for several nominally identical cables of length 30 m, and for several cables of length 30 cm, for three different diode arrays, each measured multiple times. The temperature uncertainty is determined from the estimated difference in temperature between diodes in the array (2°C uncertainty) and the thermocouple temperature reading on the PC board (0.3°C uncertainty). The temperature uncertainty is determined to be 17.8% and includes leakage current temperature scaling according to [5]:

$$I(T) = I(T_R)/R(T), \text{ where}$$
$$R(T) = (T_R/T)^2 \exp\left(-\frac{E_{\text{eff}}}{2k_B}(1/T_R - 1/T)\right)$$

with a reference temperature $T_R = 25$ °C. The standard deviation of three measurements of leakage current per diode is collected, producing a statistical uncertainty given by the standard deviation on all diodes in the array, 0.4%. The total uncertainty is the quadrature sum of the four contributors.

**The Cooling System**

Cooling is provided to several points in the instrument. A Norhof liquid nitrogen (LN2) microdose dewar system is positioned externally to the radiation enclosure. Techflex vacuum-insulated stainless steel transfer lines of inner diameter 6.4 mm extend 30 meters from the dewar into the radiation enclosure where they adapt to the sample boxes. Optionally, to limit activation of transfer line material within a meter of the beamline, the stainless tubing can be interfaced with conventional plastic tubing insulated with Aerocel (EPDM) foam over the last few feet before the sample boxes. While the sample box construction is not designed to be hermetic, this nitrogen vapor inhibits condensation on the samples which may be powered.

The cooling system uses LN2 at -196 °C as a cooling medium. A microprocessor-controlled pump on the dewar is attached to a heating element. The element heats the LN2 slightly, producing an overpressure in the dewar that drives LN2 into the transfer line. The system uses Proportional Integral Derivative (PID) temperature control. Figure 9 shows the elements of this system. The set-point is the desired temperature of the sample box. The Resistance Temperature Detector (RTD) measures the temperature inside the sample box and compares that with the set point. The difference between the two values is the Error_Value. This value is used in calculating Proportional, Integral, and Derivative terms which produce a signal which either increases or decreases the LN2 flow by controlling the heater. The RTD measures the new temperature (the New_Feedback_Value) and this feedback loop iterates.

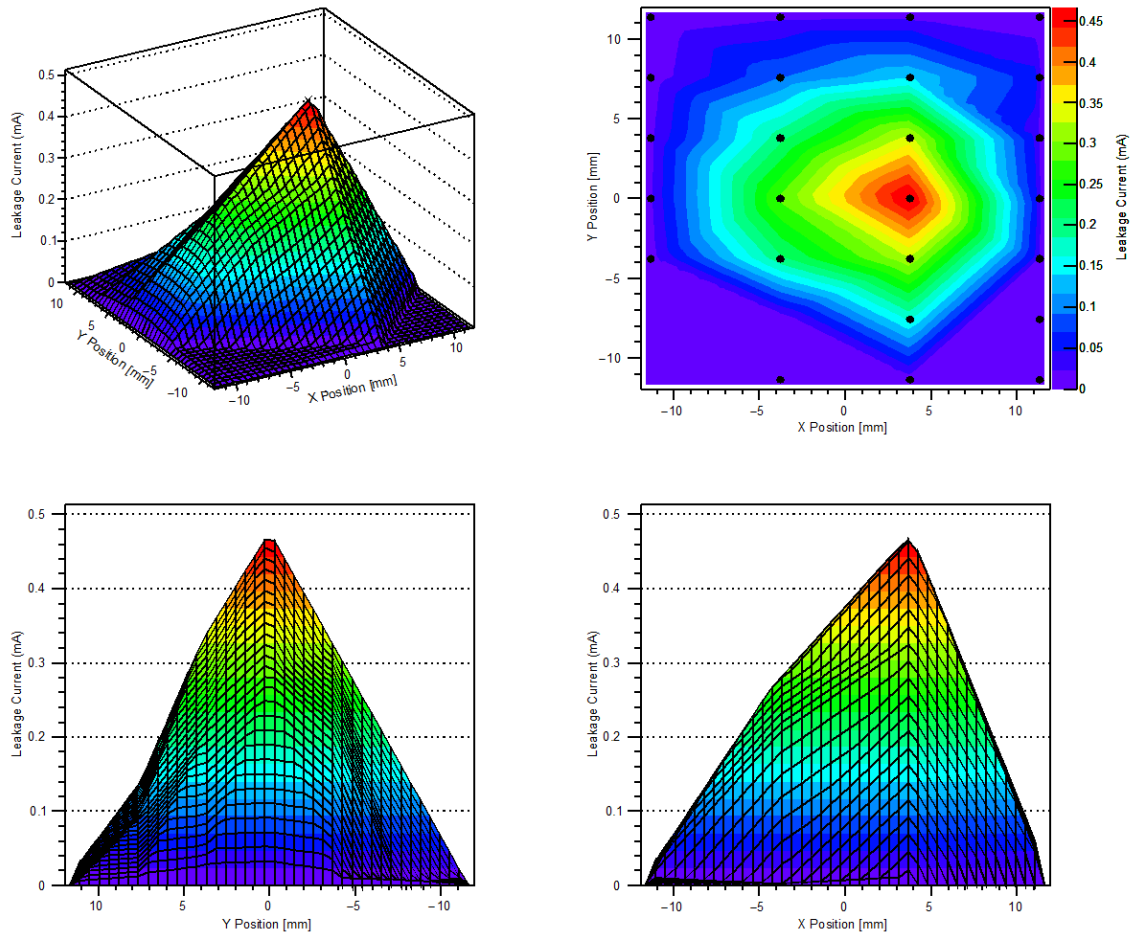

Figure 8: The 800 MeV proton beam profile, imaged using the 3D diode array. (Lower left) The recorded leakage current versus position, projected in the y-direction; (Lower right) The recorded leakage current versus position, projected in the x-direction; (Upper left) three-dimensional reconstruction of the leakage current recorded as a function of x- and y-positions; (Upper right) contour plot of leakage current in the x-y plane, with black dots indicating the locations of the centers of the active diodes.

| Contributors to the uncertainty | Packaging | Readout cable | Temperature | Statistics | Total |
|---|---|---|---|---|---|
| Uncertainty (%) | 8.2 | 11.7 | 17.8 | 0.4 | 22.8 |

Table 1. Sources of uncertainty in the 3D diode array measurement.

The rate at which the Norhof system equilibrates the samples to the desired temperature depends upon the P, I, and D parameters, two time constants (Td_S, the delay time constant during steady state operation, and Td_B, the delay time constant during boost), and the height of the maximum point of the fill line to the exhaust valve at the dewar. We have operated this 30-meter line in an experimental hall of nominal temperature 25°C and, for a single cooling branch, lowered the temperature in a sample box of a few liters' volume to -10°C and maintained the atmosphere at that temperature indefinitely. Table 2 shows a sample set of the Norhof parameters that when applied with a transfer line of length 30 m, achieve -10°C at the destination in 140 minutes.

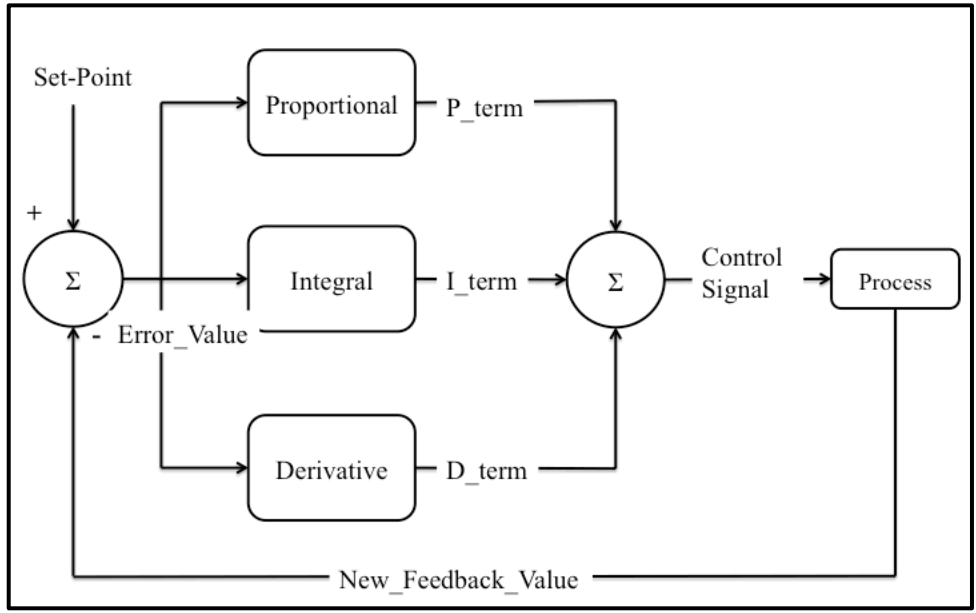

Figure 9: The PID temperature control system.

The Norhof drip rate varies from a few drops per minute to 1 liter per minute throughout the cooling process in order to reach the set temperature and maintain a constant temperature afterwards. During operation, the temperature measurements by the RTD in the Norhof system are validated by a thermocouple connected externally. Figure 10 shows both measurements as a function of time, for a trial that required a sample temperature of -5°C.

| Parameter | Value |
|---|---|
| Proportional | 9 |
| Integral | 8 |
| Derivative | 8 |
| Td_B (seconds) | 7 |
| Td_S (seconds) | 8 |
| Fill line height (cm) | 76 |

Table 2. Norhof parameters applied to achieve a -10°C atmosphere in a few-liter sample box served by a 30 m insulated transfer hose, in a nominally 25°C experimental hall.

**Operational Experience and Conclusions**

Several versions of this instrument have been operated successfully in the 800 MeV proton beam at the Los Alamos Neutron Science Center during roughly annual radiation campaigns from 2014 to the present; each exposure lasted approximately 48 to 72 hours.

All elements of the instrument work reliably. Some images of the instrument in situ at LANSCE are shown in Figure 11.

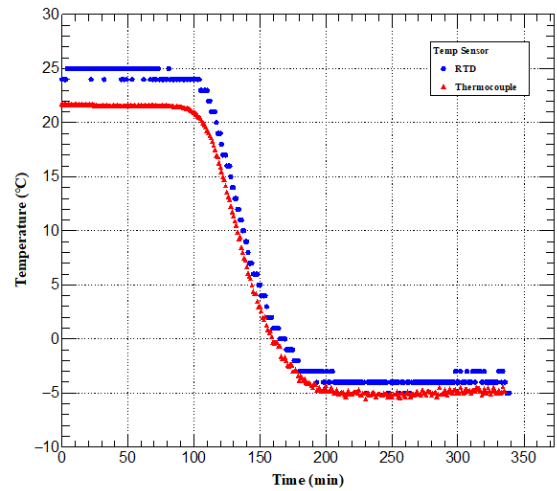

Figure 10: Temperature recorded in the sample box located 30 m from the Norhof dewar, as a function of time after the cooling process was started. Temperature was measured with two sensors, a thermocouple and an RTD.

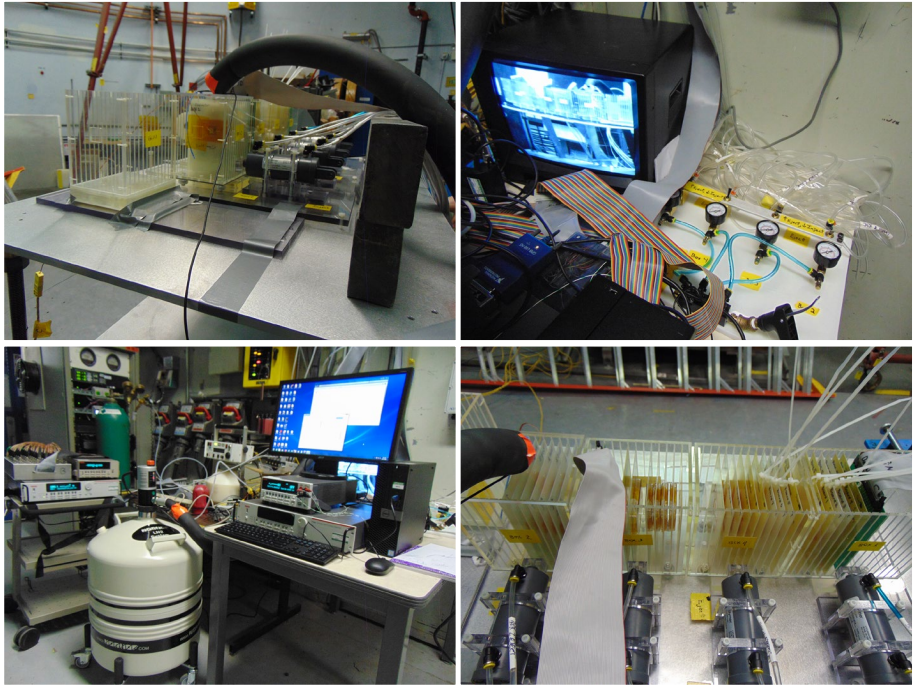

Figure 11. A composite of images of the instrument, including prototypes, photographed over several years during irradiation campaigns at LANSCE.

Both the p-i-n diode array and the 3D diode array are effective tools for measuring particle beam profile and fluence. The 3D diodes have the benefit that they can operate above fluences on the order of $1.7 \times 10^{16}$ 1-MeV-neutron-equivalent/cm$^2$ ($n_{eq}$). This fluence, and those recorded by p-i-n diode arrays, are confirmed by post-irradiation gamma spectrometry of ultrapure aluminum foils. These segmented foils are placed directly atop 3D arrays during each run, and thus each foil receives exactly the same fluence as the diodes beneath it.

The p-i-n diode arrays, on the other hand, must be replaced after each approximately $4 \times 10^{15}$ $n_{eq}$. As is described in Ref. [1], above fluences of $4 \times 10^{15}$ $n_{eq}$, the linear relationship between applied fluence and measured forwarded voltage ceases to hold for the p-i-n diodes.

Drawbacks of using the 3D arrays include the facts that (1) they are photosensitive, which necessitates radiation resistant light shielding directly in the path of the particle beam, (2) they require cooling to suppress leakage current as radiation and annealing progress, and (3) conversion of their signal to an absolute fluence requires an accurate thermal history for them, which may be difficult to assure in the conditions of the active beamline.

While all of these drawbacks to the 3D implementation could be overcome through advanced packaging, the p-i-n diode array suffers from none of them, and its lower radiation tolerance can be overcome by sequential use of multiple p-i-n arrays inserted and removed remotely via the translation system described here. For example, during a 48-hour campaign, if the target fluence is $10^{16}$ $n_{eq}$, three sets of p-i-n arrays need to be cycled sequentially into and then out of the beamline.

The cooling system is able to operate over substantial distances to achieve and maintain temperatures as low as those typically used in particle physics experiments. Limitations on the circulation pattern of coolant in the sample boxes might be overcome by incorporation of a vacuum port at the opposite end of the box from the coolant inflow.

**Acknowledgments**

This work was supported by the National Science Foundation Major Research Instrument program through award 1623479. Dr. Gian-Franco Dalla Betta of the Department of Electrical Engineering, University of Trento, and Dr. Maurizio Boscardin of Fondation Bruno Kessler provided bare 3D diodes which were used by the authors in the design and construction of the monitoring arrays and of their readout circuit. Carl Allen, of Allen Instrument Co. Inc., proposed ingenious solutions to some design challenges. Use of the 800 MeV proton beam at the Los Alamos National Laboratory permitted critical testing of prototypes and validation of the final device; the authors are especially grateful to the instrument scientists (Dr. Michael Mocko, Dr. Ron Nelson, Dr. Stephen Wender, and Dr. Leo Bitteker) and to the accelerator operators.